\documentstyle[aps,epsf,twocolumn]{revtex}   
\begin{document}   
 
\title{Higher Order Effects in the Dielectric Constant of \\
Percolative Metal-Insulator Systems above the Critical Point } 
  
\author{W.D. Heiss, D.S. McLachlan and C. Chiteme}

\address{Department of Physics\\
 University of the Witwatersrand,
PO Wits 2050, Johannesburg, South Africa}

\maketitle


\begin{abstract} 
The dielectric constant of a conductor--insulator mixture
shows a pronounced maximum above the critical volume concentration.
Further experimental evidence is presented
as well as a theoretical consideration based on a phenomenological
equation. Explicit expressions are given for the position of the maximum in
terms of scaling parameters and the (complex) conductances of the conductor
and insulator. In order to fit some of the data, a volume fraction dependent
expression for the conductivity of the more highly conductive component is
introduced.
\end{abstract}  
\def\si{\sigma _I}
\def\sm{\sigma _M}
\def\sc{\sigma _C}
\def\smc{\sigma _{MC}}
\def\d{{\rm d}}

The ac and dc conductivities of resistor and resistor-capacitor (RC)
networks and continuum conductor-insulator composites have been extensively
studied for many years. In systems where there is a very sharp change
(metal-insulator transition or MIT) in the dc conductivity at a critical
volume fraction or percolation threshold denoted by $\phi _c$, the most
successful models, for both the dc and ac properties, have proved to be
approaches using 
percolation theory. Early work concentrated on the dc properties, but since
it was realized that the percolation threshold is a critical point, and that
the percolation equations could be arrived at from a scaling relation,
several papers (\cite{jun} and the references therein), reporting experimental
results on the ac conductivity have appeared. Review articles, containing
the theory and some experimental results, on the complex ac conductivity and
other properties of binary metal-insulator systems include \cite{jpcl}-
\cite{cew}.

In \cite{jun}, \cite{wumcl} and \cite{dsm}
the following equation was introduced

\begin{equation}
{(1-\phi )(\si ^{1/s}-\sm ^{1/s})\over
(\si ^{1/s}+A\sm ^{1/s})}+
{\phi (\sc ^{1/t}-\sm ^{1/t})\over
(\sc ^{1/t}+A\sm ^{1/t})}=0,
\label{gem}
\end{equation} 
which gives a phenomenological relationship between $\sc ,\si ,$
and $\sm $. They are, respectively, the conductivities of the conducting and
insulating component and the mixture of the two components. Note that all
three quantities $\sc ,\si ,$ and $\sm $ can be real or
complex numbers in Eq.(1). The conducting volume fraction $\phi $ ranges
between 0 and 1 with $\phi =0$ characterizing the pure insulator substance $ 
\left( \sm \equiv \si \right) $ and $\phi =1$ the pure conductor
substance $\left( \sm \equiv \sc \right) $. The critical volume
fraction, or percolation threshold is denoted by $\phi _c$, where a
transition from an essentially insulating to an essentially conducting
medium takes place, and $A=\left( 1-\phi _c\right) /\phi _c$. For $s=t=1$
the equation is equivalent to the Bruggeman symmetric media equation 
\cite{dsmbl}.
Equation (1) yields the two limits

\begin{equation}
\label{limm}
|\sc |\to \infty :\qquad
\sm =\si {\phi _c^s\over (\phi _c-\phi)^s} \qquad \phi<\phi _c\\
\end{equation}
\begin{equation}
\label{limp}
|\si |\to  0:\qquad
\sm = \sc {(\phi -\phi _c)^t\over (1-\phi _c)^t} \qquad \phi >\phi _c
\end{equation} 
which characterize the exponents $s$ and $t$. Note that Eqs. (2) and (3) are the
normalized percolation equations. At $\phi =\phi _c$,

\begin{equation}
\smc = \sc /A^{\left( st\right) /\left( s+t\right)
} \left( \si /\sc \right) ^{t/\left( s+t\right)
}
\label{atfc}
\end{equation}
up to higher order terms in $\si /\sc$.

Thus, in the crossover region where $\phi $ lies between $\phi _c-\left( \si
/\sc \right) ^{1/\left( s+t\right) }$ and $\phi _c+\left( \si
/\sc \right) ^{1/(s+t)}$  we have \cite{jpcl}-\cite{cew}

\begin{equation} 
\sm \propto \si ^{t/\left( s+t\right) }\sc ^{s/\left(
s+t\right) }.
\label{cross}
\end{equation}

When using the above equation to analyze ac systems the complex
conductivities $( \sigma _x=\sigma _{xr}-i\omega \epsilon _o\epsilon
_{rx}$, where $\omega $ is the angular frequency, $\epsilon _o$
the permittivity of free space and $\epsilon _{rx}$ the real
relative dielectric constant of the component) must be inserted in
Eq.(\ref{gem}). Below $\phi _c$ the leading term yields the imaginary 
conductivity or real dielectric constant, while the next
order term gives the real conductivity. In turn, 
above $\phi _c$ the real conductivity is the leading
term and the real dielectric constant appears in the next order term, 
upon which the present paper focuses.
In many instances the approximation $\sc =\sigma _{cr}$ and $\si
=-i\omega \epsilon _o\epsilon _r$ is made \cite{jpcl}-\cite{cew}. In 
this case, the crossover
region ranges over $\phi $ between $\phi _c-\left( \omega \epsilon
_o\epsilon _r/\sc \right) $ and $\phi _c+\left( \omega \epsilon
_o\epsilon _r/\sc \right) $. This means that, as a function of
frequency, a sample with $\phi $ close to $\phi _c$ will enter an experimentally
observable crossover region for sufficiently large $\omega $ and 
eventually obey Eq.
(5) with a frequency dependence of $\omega ^{t/\left( s+t\right) }$ for both the
real and imaginary parts.

In a series of papers reporting on the dc \cite{wumcl} and ac 
\cite{jun,dsm} conductivities of
three Graphite-Boron Nitride system systems it was shown theoretically that
\begin{enumerate}
\item the scaling relations derived from the first order terms of 
Eq.(\ref{gem}), above
and below $\phi _c$, had the frequency dependencies required by the
classical scaling ansatz \cite{jpcl}-\cite{cew} (dc scaling is dealt 
with in \cite{dsmph});
\item the second order scaling relations, derived from Eq.(\ref{gem}), in the
crossover region, had the frequency and $\left( \phi -\phi _c\right) $
dependencies required by the classical scaling ansatz \cite{jpcl}-\cite{cew};
\item the second order scaling relations, in the region where the first order
terms obey Eqs.(2) and (3), were not in agreement with the classical
scaling ansatz \cite{dsm};
\item the loss or conductivity term in the dielectric state had a frequency
exponent of $\left( 1+t\right) /t$ and not 2 as required by the classical
ansatz \cite{jpcl}-\cite{cew}.
\end{enumerate}

Experimentally it was shown that
\begin{enumerate}
\item  the first order experimental results for the G-BN systems could be scaled
onto separate curves above and below $\phi _c$ and that these scaled
curves could be fitted by the scaling functions generated by Eq.(\ref{gem}), for
all $\phi $ and $\omega $ values \cite{wumcl}. This was also found to 
be the case by Chiteme and McLachlan \cite{cchdsm2} for 
''cellular'' percolation systems, some of
the dc properties of which are given in \cite{cchdsm1};
\item after taking the loss (conductivity) term of the dielectric 
component into account 
the exponent for the loss was much closer to $(1+t)/t$ than to
2 \cite{dsm};
\item  the real dielectric constant, when plotted as a function of $\phi $
for the fixed $\omega $, does not peak at $\phi _c$ but continues to
increase reaching a peak at a value of $\phi $ greater than $\phi _c$.
The resultant curve has a hump like appearance and the position of the peak
depends on $\omega $, $\epsilon _{ir}$ and $\sc $. It was then shown
\cite{dsm} that this behavior could be qualitatively modeled by
Eq.(\ref{gem}).
\end{enumerate}

In the present paper progress is reported on this last point. We begin with
a further theoretical investigation of the behavior of the dielectric 
constant $\epsilon $ just above $\phi _c$. 
Since $\sm $ is in general the result of
finding numerically the root of a transcendental equation, this can only be 
achieved by expanding $\sm $ around $\phi _c$. With the ansatz
$\sm (\phi )^{1/t} =\smc ^{1/t}+\delta $ we obtain from Eq.(\ref{gem})
\begin{equation}
\label{delta}
\delta ={\smc ^{1/t}\si ^{1/s} (A-1)+\Delta \phi \,\sc ^{1/t} \smc ^{1/s}
\over A(t/s+1)\smc ^{1/s}-t/s \,\Delta \phi \,\sc ^{1/t} \smc ^{1/s-1/t}
-(A-1)\si ^{1/s} }
\end{equation}
 with $\Delta \phi =(\phi -\phi _c)/\phi _c$. 
Note that $\delta $ does not vanish when $\Delta \phi =0$ as we have to take
into account additional terms of lower order which do not vanish for
$\phi =\phi _c$. These terms are omitted in Eq.(\ref{atfc}) but they are of 
the same order as terms linear in $\phi -\phi _c$ and must be incorporated
for reasons of consistency. Inserting the right hand 
side of Eq.(\ref{atfc}) for $\smc $ it is now
straightforward, albeit tedious, to obtain the position of the maximum of
the imaginary part. The calculation yields for the maximum
\begin{eqnarray}
\label{fmax}
\phi _{{\rm max}} &=& 
\phi _c+{s+t\over 2t}{(1-\phi _c)(1-2 \phi _c)\over \phi _c}A^{-{2s\over s+t}}
\left({|\si |\over \sc}\right)^{{2\over s+t}} \\  &+& 
{\cal O}(\left({\si \over \sc}\right)^{{3\over s+t}}) \nonumber .
\end{eqnarray}
We emphasize that the deviation of $\phi _{{\rm max}}$ from $\phi _c$ is 
obtained only when the expansion in powers of $(\si /\sc )^{1/(s+t)}$ is 
taken beyond the first power; in fact, the result given can be obtained only 
if the expressions are expanded up to the third power.

\begin{figure}
\epsfxsize=3.0in
\centerline{
\epsffile{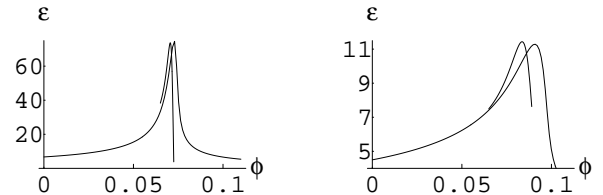}}
\vglue 0.15cm
\caption{
Comparison of the analytic expression $\sm (\phi ) $ with the numerical
solution for 10 Hz (left) and 1 Mhz (right). The curve for the analytic
$\sm (\phi )$ has been drawn only for values of $\phi $ where the
analytic expression for $\delta (\phi )$ is supposed to be reliable.
The parameters used are given in the text.}
\label{fig1}
\end{figure}

We have checked the reliability of this analytic expression by comparing
with numerical solutions over a wide range of frequencies and are satisfied
with its performance (see Figs.1). The parameters chosen in the illustration
adhere to the Niobium Carbide system \cite{dsm}
which exhibits pronounced maxima beyond
$\phi _c=0.065$; they are $t=5.46 (4.71),\, s=0.75 (0.40)$, 
$\sc =1.07\cdot 10^4 (\Omega m)^{-1}$ and
$\si = -i\omega \epsilon_0 \epsilon_r$ with $\epsilon_r= 5.4 (6.7)$ for 
the left (right) case.
The difference between the two curves, even at $\phi =\phi _c$, reflects the 
error made by considering linear terms in $\delta $ only.
Nevertheless, the analytic expression allows important
conclusions to be drawn for the position of the maximum. Note that
$\phi _{{\rm max}}$ does not depend  on individual values of $\si $ or 
$\sc $, in other
words the maximum position depends only on the ratio $\si /\sc $ (or 
equivalently on $\omega \epsilon _0 \epsilon _r/\sc $). Note in particular that
\begin{enumerate}
\item the deviation of $\phi _{{\rm max}}$ from $\phi _c$ starts with
the second order term in $(\si /\sc )^{1/(s+t)}$; recall that this first order
term determines the width of the cross over region \cite{jpcl}-\cite{cew};
\item the larger the ratio $\si /\sc$ the further is the maximum pushed away 
from the transition point $\phi _c$. In turn, the position of the
maximum tends towards $\phi _c$ for $\si /\sc \to 0$ as it should;
\item for finite values of the ratio $\si /\sc $ the distance
$\phi _{{\rm max}}-\phi _c$ increases the more rapidly the more pronounced
the inequality $s+t>2$;
\item to lowest order, the frequency dependence of of the distance
$\phi _{{\rm max}}-\phi _c$  is given by
$\omega ^{{2\over s+t}}$.
\end{enumerate}
The derivative of $\Im \sm$ at $\phi _c$ is easier to obtain, but the result
is slightly more involved. We here report the essential result
$$ {\d \Im \sm \over \d \phi }|_{\phi _c}
\propto \Im \left[\left({\si ^{{t\over s}} \over 
\sc ^{{s\over t}}}\right)^{{1\over s+t}}
\si^{{1\over s}}\sc ^{{1\over t}}\right] .$$
We stress that the derivatives of $\Re \sm $ and $\Im \sm $ at $\phi _c$ are 
always continuous and tend to zero for $(\si /\sc )\to 0$.

\begin{figure}
\epsfxsize=3.0in
\centerline{
\epsffile{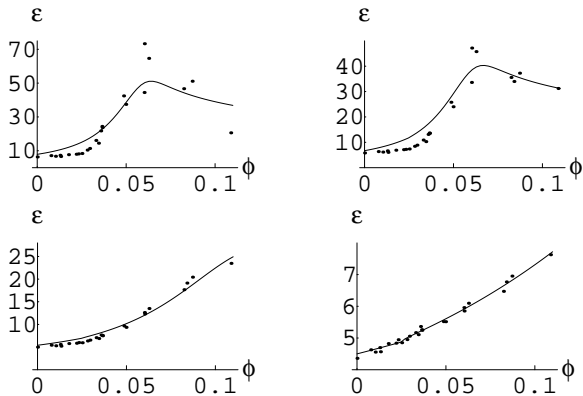}}
\vglue 0.15cm
\caption{
Fits of experimental data at 1 Hz, 10 Hz, 1 KHz and 1 MHz (top left to
bottom right) for conducting $Fe_3O_4$ grains in a wax coated
talcum powder matrix. The parameters used are given in the text.
}
\label{fig2}
\end{figure}

As it has now been established and confirmed experimentally in the
present paper and in \cite{dsm},
Eq.(\ref{gem}) gives rise to the hump in the dielectric constant just above
$\phi _c$. However, as can be seen in \cite{dsm}, Eq.(\ref{gem}), as it
stands, only qualitatively
models the experimental data. In order to better fit the dielectric data we
propose an effective dependence of $\sc $ on $\phi $ that is based in spirit
on a model due to Balberg \cite{ibalb} for non--universal values of $t$. 
The model accounts for values of $t$ higher than those allowed by the Random
Void (RV) and Inverse Random Void model (IRV) \cite{half,fengh}. In this model
Balberg assumes that the resistance distribution function $h(\epsilon )$, where
$\epsilon $ is the proximity parameter, has the form $\epsilon ^{-w}$ as
$\epsilon \to 0$ and does not tend towards a constant as in the RV and IRV 
models. Using the approach of \cite{fengh,tfeng} and keeping the underlying
node links and blobs model \cite{stauff}, Balberg derives an expression for a 
non-universal $t$ which is equal to the one obtained from the RV model when 
$w=0$ but can give a larger $t$ for $w>0$. For $w>0$ the model also shows that
the average resistance in the network can diverge when $\phi \to \phi _c$. The
increase in $t$ beyond its universal value $t_{{\rm un}}$ as found by Balberg
can be lumped into $t=t_{{\rm un}}+t_{{\rm nun}}+r$ where $r$ is the extra 
contribution due to the {\it characteristic}
resistance of the network diverging at $\phi \to \phi _c$. This
increase in {\it characteristic} resistance is incorporated into Eq.(\ref{gem})
and hence also into Eqs.(\ref{limm}) and (\ref{limp}), by substituting 
$\sc $ with

\begin{equation}
\sc ^{{\rm eff}} =\sigma _{00}+\sigma _{c0}\left( {\phi -\phi _c \over
1-\phi _c }\right)^r,\qquad \phi >\phi _c
\label{repl}
\end{equation}
with $r>0$.
The solution of Eq.(\ref{gem}) should yield a continuous $\sm $ across 
$\phi _c$, which means that the
conductivity $\sc $ should nowhere vanish. As a consequence, 
$\sc $ must reach a non--zero value denoted by $\sigma _{00}$ at 
$\phi _c$ which is expected to be considerably lower than $\sigma _{c0}$.
To avoid too many parameters the simplest
assumption was made for $\phi <\phi _c$, that is $\sc \equiv \sigma _{00}$
This should be reasonably valid just below $\phi _c$ where our interest 
is focussed. 
We stress that the effect of the modification by Eq.(\ref{repl})
is virtually indiscernible with regard to first order effects for the
solution $\sm $ of Eq.(\ref{gem}) or the corresponding percolation power laws
(Eqs. (2) and (3)) in the region where the power laws are obeyed. Computer
simulations show that using Eq.(\ref{repl}) or a constant $\sc $ 
in Eq.(\ref{gem}) causes a difference in 
$\sm (\phi -\phi _c)$ ($\phi >\phi _c$) or $\epsilon _M(\phi _c-\phi)$
($\phi _c >\phi $) which is too minute to be resolved from 
available experimental data. The same holds for dispersion plots  against 
$\omega $ of $\sm $ above $\phi _c$ and $\epsilon _M$ below $\phi _c$
as given in \cite{jun}.
For this reason it has never been necessary previously to consider the
modification given by Eq.(\ref{repl}). However, in our context
the higher order effects are discernible in the dielectric constant
just beyond $\phi _c$ 
and can be fitted much more satisfactorily than in \cite{dsm} using
Eqs.(\ref{gem}) and (\ref{repl}).

\begin{figure}
\epsfxsize=3.0in
\centerline{
\epsffile{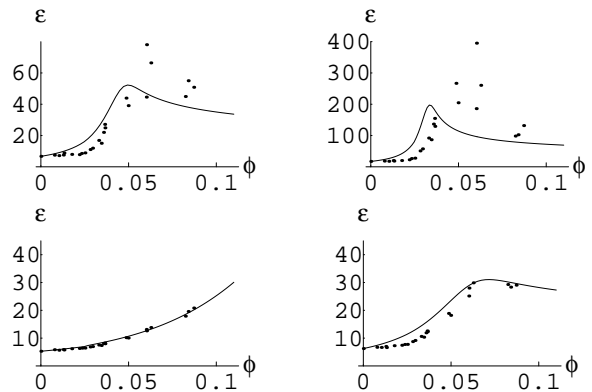}}
\vglue 0.15cm
\caption{
Fits of experimental data at 1 Hz for $T=25^0$ (top left) and $T=120^0$ 
(top right). The bottom row displays corresponding results for 1 KHz.
The system is the same as in Fig.2. The parameter values are given in the text.}
\label{fig3}
\end{figure}

In Fig.2 we display data and corresponding fits for conducting $Fe_3O_4$ 
grains in a wax coated talcum powder matrix \cite{cchdsm2,cchdsm1}. The 
fits are far superior than those presented in \cite{dsm}. We have chosen
(somewhat arbitrarily) $\sigma _{00}=\sc /10$ with the values $\sc =
\sigma _{c0} = 2.63\cdot 10^{-1} (\Omega m)^{-1}$ obtained from extrapolating
dc experimental results to $\phi =1$, and
$\si=-i\omega \epsilon _0 \epsilon _r$ where $\epsilon _r$ is measured
separately at each frequency.  Good fits are obtained for 
$(t-r,r,s)=(4.7,0.5,0.97),(4.3,0.7,0.98),
(4.4,0.6,1.0)$ and (5.6,0.4,1.6) when moving from the top left to the bottom
right display in Fig.2. The value $\phi _c=0.025$, obtained from dc 
measurements, is used in Figs.2 and 3.
Note that $t=t_{{\rm un}}+t_{{\rm nun}}+r$ is
somewhat larger than the separately measured non-universal dc value of 4.2. Also
note that in all cases it turned out 
that $r<1$. This implies a steep increase of $\sc ^{{\rm eff}}$ just above
$\phi _c$.   The fits are not necessarily optimal as the multi-parameter
landscape of the least square expression $|\sm ^{{\rm exp}}-\sm ^{{\rm num}}|^2$
in the parameters $t,s$ and $r$ (for fixed $\sigma _{00}$) has many local 
minima. However, the quality of the fits at the different local minima does not
vary greatly for acceptable fits. 

While this second order behavior of $\epsilon $ is in principle a complicated
function of the various parameters the position of the maximum at $\phi _{{\rm
max}}$ is essentially dependent only on the quotient $\si /\sc $. This
implies that a decrease of frequency (which is a decrease of $\si $) has an
effect similar to a corresponding increase of $\sc $, which can be 
obtained by an increase of temperature. This is convincingly demonstrated in 
Fig.3 where fits are obtained for the same system as in Fig.2 at different
temperatures, i.e. at different values of $\sc $ and $\epsilon _r$. The
experimental data and the theoretical curves show that an increase in $\sc $
is equivalent to a decrease in $\omega $ (or $\si $). The experimental 
data appear somewhat erratic as they usually do for the low frequency chosen 
(1 Hz). The data in the first column of Fig.3 are slightly different from those
in the first column of Fig.2 as the former are taken in a temperature
controlled oven. Much better fits are obtained for higher
frequencies as illustrated in the bottom row of Fig.3 (1 KHz) as the 
dielectric data are more reliable at higher frequencies. 
Similar to Fig.2, experimental values are used for
$\epsilon _r(\omega , T)$ with $\sigma _{c0}=\sc =3.22\cdot 
10 ^{-1}(\Omega m)^{-1}$ at $25^0C$ and $\sigma _{c0}=\sc =1.42 
(\Omega m)^{-1}$ at $120^0C$. 
The fits were obtained with the parameters
$(t-r,r,s)=(4.65,0.5,1.0),(3.4,0.6,1.0),(4.4,0.6,1.0),(3.5,0.9,1.0)$ again
choosing the fixed value $\sigma _{00}=\sc /10$.

\begin{figure}
\epsfxsize=3.0in
\centerline{
\epsffile{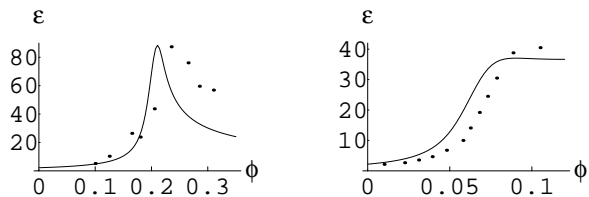}}
\vglue 0.15cm
\caption{
Fits of experimental data at 100 KHz of a water--oil emulsion using a hydrophil
agent for uniform droplet size.}
\label{fig4}
\end{figure}

In continuum systems, all models for a non universal $t$ are based 
on the premise \cite{ibalb}-\cite{fengh}
that a distribution of conductances exists in the conducting component which 
has a power law singularity. In a system where no such 
singularity occurs it should be possible to fit the data using Eq.(1), 
with $t$ values closer to 2 and without the use of Eq.(8). A system {\it par 
excellence} which satisfies this requirement is a water-oil emulsion, using 
a surfactant to ensure uniform drop size. Data for such systems has been 
published in [16] -[18] and we present in Fig.4 data from Fig.1 of
[16] and from Fig.4 of [17], which have been fitted using Eq.(1) with the 
same value of $\sc $ above and below $\phi _c$. The values used, i.e. 
$\sc =1.9(1.5)(\Omega m)^{-1},\, \si =-i\,2\cdot 10^{-11} \omega ,\,\phi _c=
0.2(0.06)$ for the left (right) display and $\nu=10^5$Hz are based on the 
dielectric and conductivity curves given in the quoted papers.
The fitting curves use $s=1.15(1.35)$ and $t=1.85(2.1)$.
Note that the values of $t$ are close to the universal value. Although better 
fits should be obtained with a more precise knowledge of the system 
parameters, these results show that using Eq.(8) as input in 
Eq.(1) is necessary to fit the dielectric data only when abnormally high 
values of $t$ are 
observed in the dc data. Note that these data have been previously analyzed 
using Eq.(1) with a single exponent $(s=t)$ in [19].

To summarize: the pronounced hump observed experimentally for the dielectric 
constant of percolative metal insulator systems just above the critical 
concentration can be modeled  using the higher order terms of Eq.(1) 
near $\phi _c$. In the cases where $t$ is well above the universal value 
Eq.(8) must be used to quantitatively model the dielectric data. 
Note also that all previous experiments on the dielectric constant "below" 
$\phi _c$, where $\phi _c$ has 
not been independently measured by dc conductivity measurements, have 
probably incorrectly identified values of $\phi _c$.

\end{document}